# Title

Replication and study of the colouration of Edmond Becquerel's photochromatic images

# Authors list

Victor de Seauve[a,b], Marie-Angélique Languille[a], Saskia Vanpeene[a], Christine Andraud[a], Chantal Garnier[a], Bertrand Lavédrine[a].

[a] Centre de Recherche sur la Conservation (CRC), Muséum national d'Histoire naturelle, CNRS, Ministère de la Culture, 36 rue Geoffroy Saint Hilaire, 75005 Paris, France.

[b] SACRe (EA 7410), Ecole normale supérieure, Université PSL, 75005 Paris, France.

# Abstract

Edmond Becquerel's process of making photochromatic images, which are the first colours photographs, is still poorly understood. In particular, the origin of the colours of these images gave rise to a long-lasting debate, starting from the publication of Becquerel in 1848 until the end of the XIX[th] century. This photographic process was replicated in the laboratory, and the colouration of the sensitized plates was studied by means of visible spectroscopy. A comprehensive description of the sensitization routes that were adapted from Becquerel's writing is given. The study of the exposition step allowed us to gain insights in the colouration mechanisms of the sensitized plates, and to compare between the two sensitization routes developed by Becquerel in terms of spectral sensitivity and colour faithfulness. The so-called "electrochemically sensitized" samples were found to be more sensitive and to reach a larger chromatic space than the "immersion-sensitized" ones. Finally, a reproducible way of creating coloured samples is suggested.

# Keyword

# License

# Highlights

- Both the electrochemical and immersion sensitization routes of Edmond Becquerel's process were successfully reproduced in the laboratory
- The exposition to visible light creates a reflectance increase at the exposition wavelength
- At a radiant exposure of 10 kJ m$^{-2}$, colours are unique and resemble that of the exposition wavelength
- Electrochemically sensitized plates are more sensitive and allow to reach a larger chromatic space than immersion-sensitized plates

# Introduction

Since the invention and the wide spread of photography during the first half of the XIX$^{th}$ century, physicians and photographs struggled with what was often referred to as "the great problem" of colour in photography [1] (all the translations are made by the authors). In 1848, Edmond Becquerel introduced the very first colour photographic process and coined the term "Photochromatic images" [2], [3]. Basing himself on Seebeck [4] (and see [5] for translation and interpretation) and Herschel's [6] works on silver chloride, he decided "to obtain silver chloride by all means" [3]. To make a photochromatic image, a pinkish sensitized layer is produced onto a polished silver plate – similar to those used for the daguerreotypes – by immersing it in a copper chloride solution [3] or by an electrochemical route, using an hydrochloric acid solution [7]. The sensitized layer is then coloured by the exposition to light and, though there is a strong component of the initial colour of the sensitized plate, the result reminds the colour of the incident light : "a photographic imprint of the solar spectrum with colours that remind its own colours […] the prismatic red gives a red hue, the yellow a yellow hue, the green a green hue, the blue a blue hue and in some cases, the white a white hue, etc." [3]. The first photochromatic images he produced were reproductions of the solar spectrum, as figure 1 shows. He then produced contact prints and took pictures in a *camera obscura* [7]. If the photochromatic images are the first responses to the problem of colour in photography, the long duration of exposure and the inability to fix the colours kept the Becquerel process from spreading and being widely used [8]. Here, we study for the first time the optical properties of the photochromatic images which dragged so much attention during the second part of the XIX[th] century by measuring samples produced in the laboratory. This study is also an opportunity to establish a protocol for the production of coloured samples that can be examined to investigate the origin of the colours of the photochromatic images.

Previously, the daguerreotypist Mark Kereun successfully replicated a process that was adapted from Edmond Becquerel by Abel Niépce de Saint-Victor [9]. The "heliochromies" he produced show distinct and recognizable colours [10]. Note that David Burder was also successful in replicating Niépce de Saint-Victor's process [11]. No visible spectroscopy analysis of these heliochromies have yet been reported. Nonetheless, the exposition of photolyzed pure silver chloride crystals and of thin Ag-AgCl films – the closest systems from the photochromatic images we found – to light have been studied. First Hilsch and Pohl [12], then Brown and Wainfan [13] and lastly Moser *et al.* [14] studied the "discoloration" or "bleaching" of darkened, photolyzed silver chloride crystals, under the influence of visible radiations of varying wavelengths. The result is a lowering of the absorbance of the crystals at the exposition wavelength. Likewise, the exposition to visible radiations of thin Ag-AgCl films, studied by Ageev *et al.*, manifests itself by a decrease of the absorbance at the exposition wavelength [15].

First, the replication in the laboratory of two sensitization routes described by Becquerel and the optical properties of the sensitized plates are presented. As details and parameters in the preparation of the images can be critical, a comprehensive description of the process is given in the Material and methods section. Then the monitoring of the exposition step to light of visible wavelengths by means of UV-Visible spectroscopy in reflectance configuration is developed. Investigating the changes in the reflection properties of the plates allowed to better understand the colours that were produced, and to compare the spectral sensitivity and the colour faithfulness of both sensitization routes. Finally, the results are commented against Becquerel's writings.

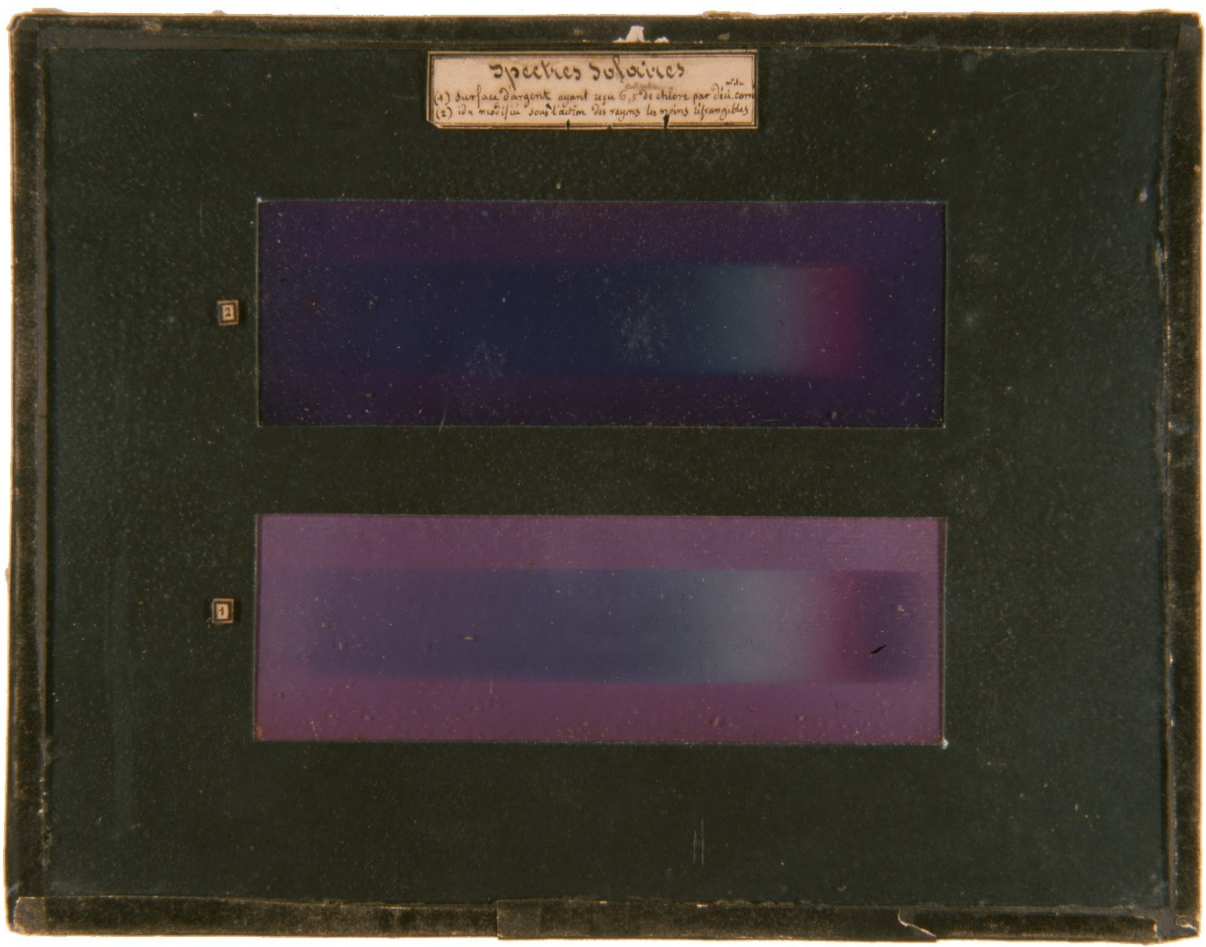

**Figure 1.** Becquerel, Edmond. Spectres solaires. 1848. Photochromatic images. Musée Nicéphore Niépce, Chalon-sur-Saône. Both solar spectra were obtained on electrochemically prepared plates, the bottom one was preexposed to infrared.

## Material and methods

Crucial details for the process replication essential to reproduce and understand the making of the photochromatic images are described hereby. Further information can be found in the experimental section.

Silver plate preparation. ¼ μm-diamond polished 30 μm-silver coated 20 × 20 mm² brass plates were chosen. X-ray Fluorescence Spectrometry (XRF) analysis and Scanning Electron Microscopy (SEM) images of the silvered plates in the supplementary information (SI) section show minor elements in addition to Ag (figure S-1), polishing trails (figure S-2) and a flat Ag layer on the brass support (figure S-3). The sensitization and the exposition steps then take place in a room where the ambient light irradiance does not exceed $1\times10^{-2}$ W m$^{-2}$ without any ultraviolet (UV) radiations.

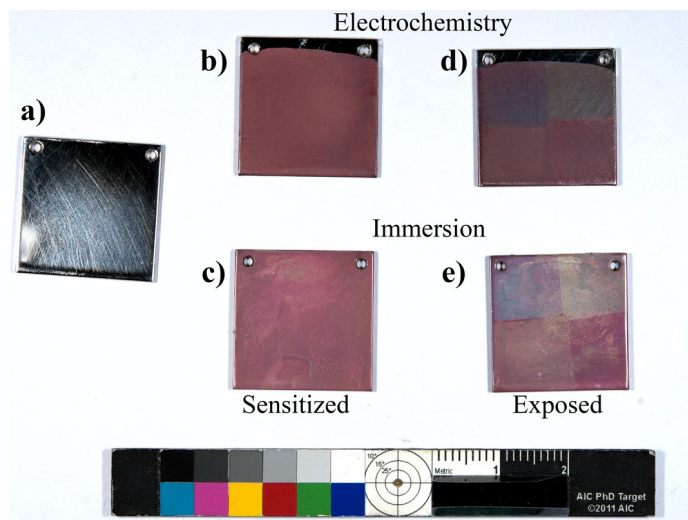

**Figure 2.** Five 20 × 20 mm² photochromatic image samples after the three steps of the process. a) Polished and degreased silver plate. b) Electrochemically sensitized silver plate after heating; the part of the sample that was immersed in HCl took a pinkish hue. c) Immersion-sensitized silver plate after heating; the whole sample was soaked in the sensitization solution. d and e) Exposed electrochemically (d) and immersion-sensitized (e) silver plate; the sensitized plates were exposed to light spots of various wavelengths, centred around 446 nm, 503 nm, 584 nm, 696 nm (from left to right, top to bottom).

Immersion sensitization. The sensitization solution was prepared according to [7] by dissolving 0.2 mol L$^{-1}$ CuSO$_4$ and 2.5 mol L$^{-1}$ NaCl in distilled water. The prepared silver plates are then soaked in the solution for 1 min, abundantly rinsed with distilled water, dried on nonabrasive and lint-free lens tissue (figure 2c). Uncarefully rinsed samples exhibit traces of sodium and copper that were detected with X-ray Photoelectron Spectroscopy (XPS); moreover, SEM images presented in the SI section show micrometric sodium chloride crystals on unrinsed sample (figure S-4 in the SI section).

Electrochemical sensitization. The prepared silver plate and a platinum grid are connected respectively to the positive and the negative poles of a ISO-TECH IPS-1820D direct current generator (cf. figure S-5 in the SI section). The electrolyte in which the plate and the grid are soaking, 2 to 3 cm apart from each other, is a 1 mol L$^{-1}$ HCl solution. The voltage is adapted between 0.3 and 0.5 V, rendering a 0.1 A current. The plates are sensitized during 50 s and then rinsed with distilled water. Making the assumption that the sensitization creates a pure

AgCl layer at the top of the silver layer, a charge transfer calculation allows us to determine the volume thickness of the sensitized layer: 3.9 μm for a 17 × 20 mm² sensitized surface (figure 2b). This order of magnitude is consistent with Becquerel's calculation: he finds an optimal sensitized layer thickness of 1.7 μm, assuming the sensitized layer is pure silver chloride.

<u>Heating of the sensitized plates.</u> Both immersion- and electrochemically sensitized plates are heated up to with a heat gun (Ryobi EHG2020LCD) set at 130 °C and 500 L min$^{-1}$.

<u>Optical properties characterization and monitoring of the exposition step.</u> The reflectance spectra of sensitized plates are first measured with a UV-visible spectrophotometer. The plates are then exposed to light of various wavelength (see the transmission spectra of the interferential filters that were used in figure 3b) and their reflectance spectra are regularly measured throughout the exposition step. The radiant exposure is calculated by multiplying the time of exposure by the irradiance of the light. Less than 0.1 % of the reflexion is specular and the shape of the diffuse and of the specular reflectance spectra are the same, which is why only the total reflectance spectra, ie. the sum of the specular reflectance and diffuse reflectance spectra, are plotted. The measurement has a very limited effect on the reflectance spectrum of the samples in regard to the effect of the exposition to visible light; indeed there is a factor of 1000 between the irradiances. The change of reflectance, ΔR, calculated as R(t) - R(sensitized), were also plotted to better assess those changes. In order to monitor the colour change of the samples, their colorimetric coordinates in the L*a*b* CIE 1976 [16] space have been calculated. For the sake of comparison, every L*a*b* coordinates series are shifted so that the sensitized sample coordinates are the same as the average coordinates of the sensitized samples.

# Results

<u>Optical properties of the sensitized plates.</u> The average reflectance spectra of immersion and electrochemically sensitized samples are displayed in figure 3a. The increase of the reflectance at the long wavelength side of the spectrum and the peak located bellow 400 nm account for the pinkish hue of the sensitized plates. The average standard deviation of the reflectance on the [200 ; 800] nm range is 3.3 % for the immersion sensitized samples and 2.5 % for the electrochemically sensitized samples; this dispersion is mainly attributed to the variability in the heating step – distance between the heat gun and the plate – and to the variability in the distance between the plates at the platinum grid in the case of the electrochemical sensitization. Spectra of samples obtained from silver platelets showing various initial surface states showed that the rugosity of the initial silver plate has no effect on the reflectance properties of the sensitized sample.

The calculated L*a*b* coordinates averages of the immersion- and the electrochemically sensitized samples are respectively (50.2; 15.5; 2.8) and (50.7; 15.8; 2.7). The colour difference $\Delta E_{76}$ between those two initial states is 1.4. More than the colour, inhomogeneities on the immersion-sensitized samples allow one to make the difference between both types of samples (figure 2) which is consistent with naked eye observation: the colour alone does not allow the distinction between an electrochemically and an immersion-sensitized sample.

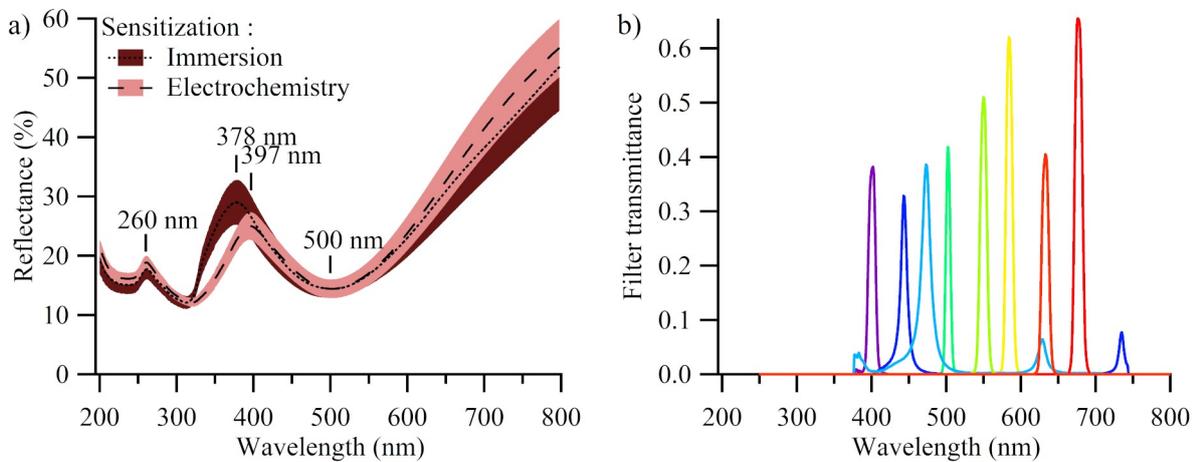

**Figure 3.** a) Average UV-visible reflectance spectra of 10 immersion and 10 electrochemically sensitized samples. The broad lines correspond to the dispersion of the data, taken as their standard deviations. b) transmittance spectra of the interferential filters used for the exposition.

Evolution of the reflection properties during the exposition. Figures 4a and b show two series of consecutive reflectance spectra, one of an immersion-sensitized sample exposed to 677 nm-centred light (a) and one of an electrochemically sensitized sample exposed to 502 nm-centred light (b). Figures 4c and d show the associated series of consecutive changes of reflectance spectra $\Delta R$.

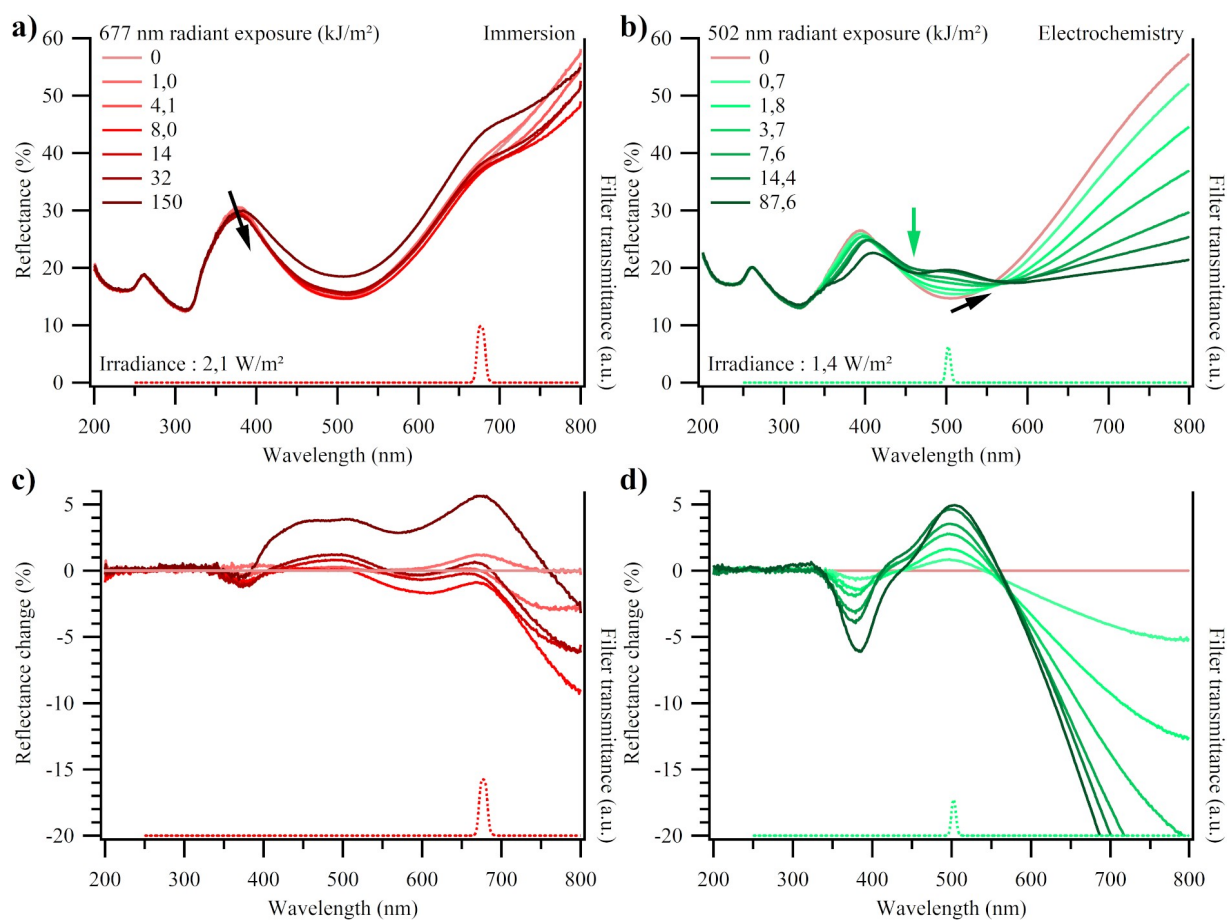

**Figure 4.** Successive reflectance spectra (and change of reflectance spectra) obtained for one immersion-sensitized sample exposed to 677 nm-centred light (a and c) and one electrochemically sensitized sample exposed to 502 nm-centred light (b and d) together with the transmittance spectra of the interferential filters (dotted lines).

During the exposition to light, the reflectance spectra undergo three major changes.

- 🎬 A decrease and a shift toward longer wavelength of the local maximum in the violet region, initially located at 378 nm for the immersion samples (black arrow on figure 4a) and at 397 nm for the electrochemical samples.
- 🎬 An increase and a shift toward longer wavelength of the local minimum in the visible region, initially located at 500 nm (black arrow on figure 4b). This is accompanied by the appearance of a second local minimum at a wavelength shorter than the exposition wavelength if the latter is shorter or equal to 500 nm (green arrow on figure 4b). If the exposition wavelength is longer than 500 nm, the second local minimum of reflectance appears at a wavelength longer than the exposition wavelength.

- 🎬 A decrease (figure 4b), followed by an increase (figure 4a), of the reflectance in the long wavelength side of the visible spectrum.

These changes can also be appreciated when looking at the consecutive changes of reflectance spectra plotted in figures 4c and d. Indeed, a dip located bellow 400 nm and a peak located around 500 nm appear in the reflectance change spectra, and their long wavelength tails decrease.

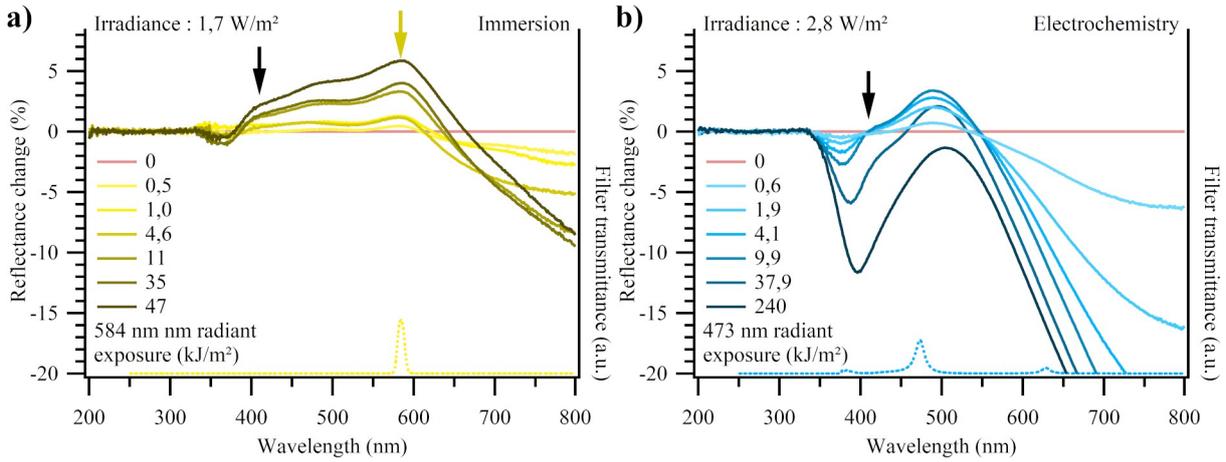

**Figure 5.** Successive reflectance change spectra obtained for one electrochemically sensitized sample exposed to 584 nm-centred light (a) and one immersion-sensitized egxposed to 473 nm-centred light (b) together with the transmittance spectra of the interferential filters.

Figures 5a and b show two series of consecutive changes of reflectance spectra, of an immersion-sensitized sample exposed to 584 nm-centred light (a) and of an electrochemically sensitized sample exposed to 473 nm-centred light (b). In addition to the above-mentioned changes, the change of reflectance spectra exhibits a shoulder around 410 nm for all the exposition wavelengths (black arrows on figures 5a and b). And for all the exposition wavelengths longer than 502 nm, another shoulder centred on the exposition wavelength appears during the exposition (yellow arrow on figure 5a). The reflectance change at the exposition wavelength can be positive or negative; so the shoulder indicates either that the reflectance has more increased, or that is has less decreased at the exposition wavelength. In a later stage of the exposition step, the reflectance spectra seem to undergo no further change or to undergo a general decrease of the reflectance with a lower efficiency than during the first times of the exposition.

Evolution of the colour during the exposition. Figures 6a and b show the evolution of the lightness L* during the exposition to light of visible wavelength versus the radiant exposure for immersion- (a) and electrochemically (b) sensitized samples. Figures 6c and d show the evolution of the hue in the a*b* frame immersion- (c) and electrochemically (d) sensitized samples.

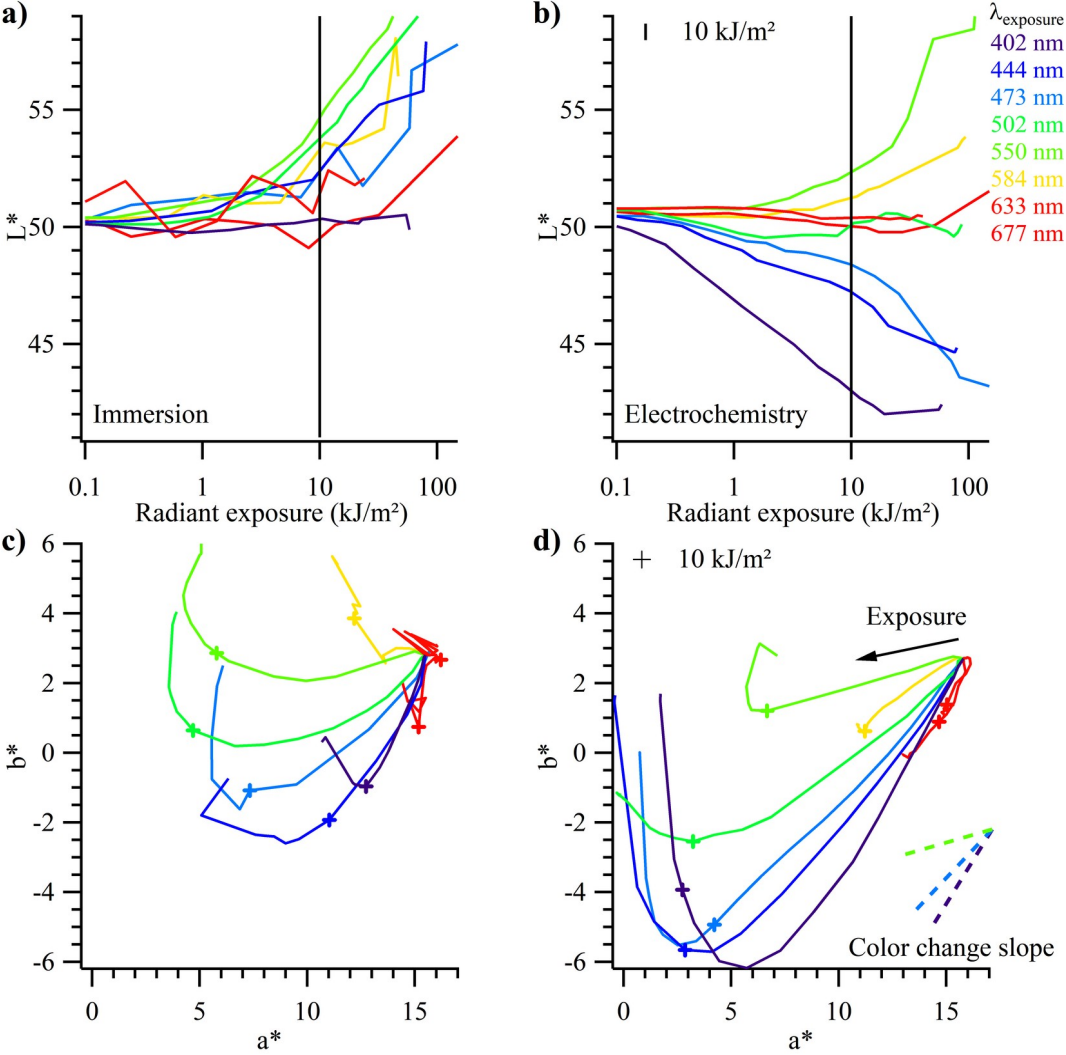

**Figure 6.** a and b) evolution of the lightness L* of the immersion- (a) and electrochemically (b) sensitized samples during their exposition to light of visible wavelengths plotted versus the radiant exposure.

c and d) evolution of the hue in the a*b* space of the immersion- (c) and electrochemically (d) sensitized samples during their exposition to light of visible wavelengths. Some examples of directions of colour change are plotted in dashed colour lines on figure 6d.

For the immersion-sensitized samples, the lightness L* remains constant or increases during the exposition to visible light with an increasing efficiency when the exposition wavelength is near to 550 nm (figure 6a). For the electrochemically sensitized samples, L* behaves the same as for the immersion-sensitized samples for all wavelength that are superior or equal to 502 nm; for the shorter wavelengths, L* decreases with an increasing efficiency as the exposition wavelength decreases (figure 6b). For both type of samples, in the first times of the exposure, the samples colours follow directions of decreasing slope when the exposition wavelength increases (see examples on figure 6d). a* and b* decrease until an exposure where a* stops decreasing and b* increases again for all the wavelengths that are inferior or equal to 584 nm (figure 6c and d).

## Discussion

Replication of Edmond Becquerel's process of creating photochromatic images. At a 10 kJ m$^{-2}$ radiant exposure, figure 6 shows that the samples have distinct colours (solid black bar on figures 6a and b and markers on figures 6c and d). Figures 2d and e present an electrochemically (d) and an immersion-sensitized (e) sample that were exposed to light spots of various wavelengths, centred around 446 nm, 503 nm, 584 nm and 696 nm. The colours blue, green, yellow and red are recognizable. Thus, the replication of Becquerel's process according to his memoirs was successful: a sensitized layer was created on a silver-cladded brass support, and colours that resemble that of the incident light were obtained on it.

Colours of the photochromatic images. When exposed to light of visible wavelengths, the optical properties of the sensitive layer of the photochromatic images evolve depending on the exposition wavelength. In particular, the shoulder in the reflectance change spectra that is observed for all the wavelengths that are longer or equal to 502 nm accounts for the resemblance of the resulting hue with the incident light. In the case of samples exposed to wavelengths shorter than 469 nm, the appearance of this shoulder is probably overlaid by the decrease of the reflectance local maximum initially located bellow 400 nm and the increase of the reflectance local minimum initially located at 500 nm. The evolution of the reflectance spectra during the exposition being characteristic of the wavelength, so is the evolution of the L*a*b* values. In that respect, the resulting colours are all unique and express the changes in the reflectance spectrum of the exposed sample.

Colour faithfulness. Table 1 show the resulting colours at an exposition of 10 kJ m$^{-2}$, framed by the initial sensitive layer colour; colours are displayed according to the RGB values that were calculated from the L*a*b* coordinates [16]. This table illustrates how unique the resulting colours are and their resemblance with the incident light. Indeed, despite the apparently strong contribution of the pinkish background of the sensitized layer, colours resemble that of the incident light of exposition.

In the a*b* space (figures 6c and d), the direction of the colour change of exposed samples is constant in the first 8 to 15 kJ m$^{-2}$ of exposition, depending on the wavelength, which is why 10 kJ m$^{-2}$ was chosen as a common radiant exposure for all the wavelengths. This direction is wavelength dependant and express the colouration of the material toward the hue of the incident light. Note that the resulting colours all lie in small (a* ; b*) ranges. They include a strong component of the sensitized layers colour, upon which a wavelength dependant colour component appears during the exposition. Contrasts between colours thus play an important role in their distinctions, which questions the perception of colours in photochromatic images. During the second step of the exposition, the colours shift toward increasing b*, with a* remaining constant, for all the exposition wavelengths. This is the sign of overexposure. Becquerel mentioned this phenomenon in his second memoir: "in general we can say that […] it is in the first moments of the action of the spectrum that the hues of the photochromatic image are the closest from these of the luminous spectrum" [7]. Figure 6a shows that the overexposure of immersion-sensitized samples brightens them; figure 6b show that overexposing electrochemically sensitized samples darkens them when the wavelength is lower than 500 nm and brightens them at 550 nm and 584 nm.

**Table 1.** Colours of the sensitive samples (background) and of the samples exposed up to 10 kJ m$^{-2}$ (colour patches) to light of various wavelengths. Colours are displayed according to RGB values. For the samples, these are calculated from the L*a*b* coordinates with the CIE formulas [16]. The exposition wavelengths colours were calculated with [17].

| λ$_{exposition}$ (nm) | 402 | 444 | 473 | 502 | 550 | 584 | 633 | 677 |
|---|---|---|---|---|---|---|---|---|
| **Immersion-sensitized** | | | | | | | | |
| **Electrochemically sensitized** | | | | | | | | |

<u>Spectral efficiency of the colour change.</u> The discussion above showed that in the first 8 to 15 kJ m$^{-2}$ of exposure, the colour changes toward that of the incident light. In this range, the spectral efficiency of the colour change can then be conflated with the chromatic sensitivity, meaning the ability of the sensitized plates to be coloured by an incident radiation. Figures 7a and b show the CIE 1976 colour change ΔE*ab during the exposure of immersion- (a) and electrochemically (b) sensitized samples to visible light. By comparing the slopes of these curves, we can see that the colour change efficiency is greater as the wavelength is closer to 500 nm for immersion-sensitized samples and is greater as the wavelength is shorter for electrochemically sensitized samples.

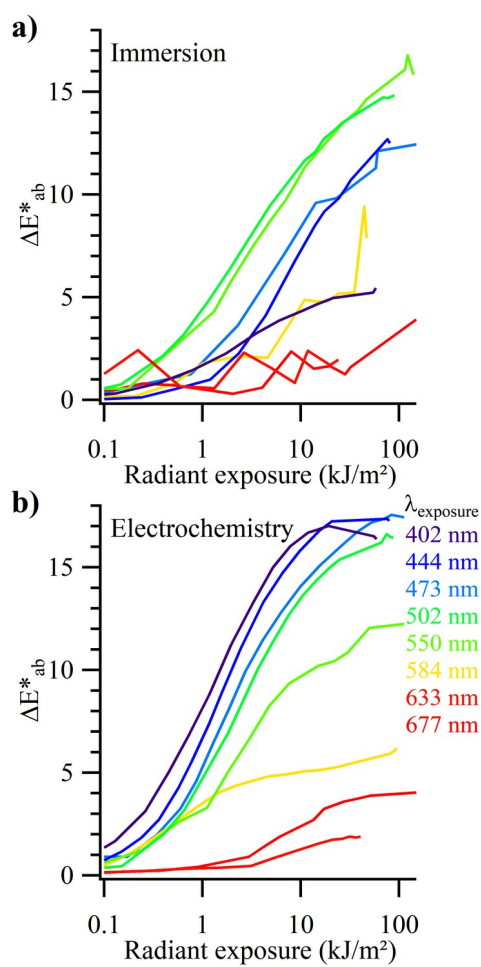

**Figure 7.** a and b) evolution of the CIE 1976 colour change $\Delta E^*_{ab}$ of the immersion- (a) and electrochemically (b) sensitized samples during their exposition to light of visible wavelengths plotted versus the radiant exposure.

Comparing figures 7a and b, we can also see that the electrochemically sensitized plates are more sensitive than the immersion-sensitized plates, especially in the blue-violet region of the visible spectra, except at 550 nm. This is apparently in contradiction with Becquerel, which states that the immersion-sensitized plates "are impressed [by the solar spectrum] more than ten times faster than any other kind of layers" [7]. Actually, Becquerel talks about the times required to get a colour that resembles that of the incident wavelength from sight, which is subjective and highly dependent on the preparation of the plate.

<u>Colour rendering of the immersion and the electrochemical sensitization routes.</u> Figures 6a and b show that for every wavelength, the lightness L* is lower for electrochemically sensitized samples than for immersion-sensitized samples. This is in agreement with Becquerel's notes, which state: "the way of operating we just described [electrochemically] offers the advantage of giving much more beautiful tints, although darker, than any other way" [7]. Markers on figures 6c and d show the a*b* coordinates of the samples that were exposed up to 10 kJ m$^{-2}$. The chroma C* – the norm of a vector in the (a*, b*) space – of immersion sensitized samples is usually higher than the C* of electrochemically sensitized samples. However, the colours spread on a wider (a*, b*) surface for electrochemically sensitized samples (d) than for immersion-sensitized samples (c). Table 2 shows the variances of the L*a*b* coordinates of theses samples, which are higher for electrochemically sensitized samples. This means that the latter can reach a wider chromatic space than immersion-sensitized samples. The colours of electrochemically sensitized samples then appear more by contrast versus the sensitized layer colour than that of immersion sensitized samples. This is consistent with what Becquerel wrote about the immersion-sensitized samples: "the hues are always somewhat grey" [7] compared to electrochemically sensitized samples.

**Table 2.** Variance of the L*a*b* coordinates of the samples exposed up to 10 kJ m$^{-2}$, depending on the sensitization method.

| Sensitization method | Variance L*$_{10\ kJ\ m^{-2}}$ | Variance a*$_{10\ kJ\ m^{-2}}$ | Variance b*$_{10\ kJ\ m^{-2}}$ |
|---|---|---|---|
| Immersion | 3.8 | 18 | 4.5 |
| Electrochemical | 9.4 | 28 | 8.8 |

# Conclusions

The two sensitization routes described by Edmond Becquerel, namely the immersion and the electrochemical, were successfully replicated in the laboratory. On both types of samples, colours were obtained and resembled that of the exposition light. The UV-visible spectroscopy monitoring of the colouration step allowed us to compare our results with Becquerel's writings and to draw conclusions on the photochromatic images colours.

The colouration is due to the lower decrease, or the higher increase, of reflectance at the exposition wavelength relatively to the initial sensitized sample. The colours of the photochromatic images are characteristic of the preparation of the sensitized layer, the exposition wavelength and the radiant exposure. At low radiant exposure, under 8 to 15 kJ m$^{-2}$, the colour change depends on the exposition wavelength; 10 kJ m$^{-2}$ was chosen as a common radiant exposure for all wavelengths in order to produce well defined coloured samples for the microstructural and chemical study of the photochromatic images. The electrochemically sensitized samples were found to be more sensitive than the immersion sensitized samples, especially in the blue-violet region of the spectrum, and to reach a larger chromatic space. The latter is in agreement with Becquerel's writings.

# Experimental

Preparation of the plates. The brass plates are silver-cladded polished with carbon black and England red by a silversmith (Orfèvrerie du Marais, 172-174 rue de Charonne – 75011 Paris, no longer in business). The plates are re-used many times: the sensitized or coloured layer is dissolved in a saturated sodium thiosulfate solution, then the plate is polished using a ¼ µm diamond suspension on a velvet polishing cloth. Before the sensitization, the plates are degreased in 96% ethanol in an ultrasonic bath for 5 minutes and then rinsed with distilled water and carefully dried on nonabrasive and lint-free lens tissue.

The platinum grid used for the electrochemical sensitization is first degreased in 35 % $HNO_3$ for 2 minutes, heated with the heat gun set on 650 °C for 2 minutes and then carefully rinsed with distilled water.

Exposition to light. The samples were exposed to light produced by a xenon arc lamp (Newport, Series Q, 75W 60056-75X-Q15) equipped with an IR and a UV filter; the condenser lens was removed in order to have a large intense and homogeneous light spot. Interferential filters centred around various wavelengths were used to produce narrow bandwidth light spots: 402 ± 5 nm, 444 ± 5 nm, 473 ± 8 nm, 592 ± 3 nm, 550 ± 5 nm, 584 ± 5 nm, 633 ± 5 nm, 677 ± 6 nm. The samples were located around 20 cm away from the light source in order to have a 5 cm diameter light spot, and to be able to measure the sample during the exposition step in *in situ* UV-visible experimental setup. The lamp was put on a camera tripod during the monitoring of the exposition step (figure 8). The samples were exposed during 20 to 30 hours and they were measured 10 to 15 times during the exposition.

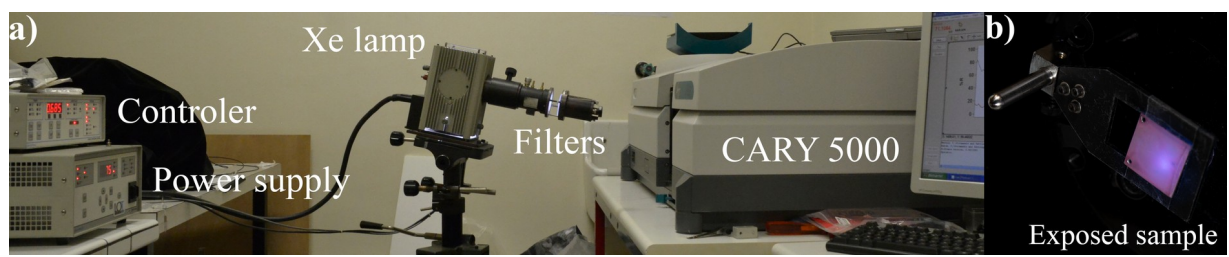

**Figure 8.** Experimental setup for the spectroscopic monitoring of the exposition step. a) Xe lamp used for the exposition. b) exposed sample.

The sample is facing inside the spectrophotometer's integrating sphere during the measurement – that is carried out with the lid closed – and facing out to the light source during the exposition. This setup allowed us to reduce the transporting of these fragile samples and the acquisition of multiple spectra on the same area.

Assessing the radiant exposure. Prior to light exposition, the irradiance was measured around the sample position with a JETI specbos 1211 spectroradiometer. 20 cm away from the light source, the usual value is around [80 – 100] W m$^{-2}$ without any interferential filter and [0.8 – 2.5] W m$^{-2}$ with the interferential filters (see figure 3b for the transmittance spectra of the interferential filters).

The radiant exposure was calculated by multiplying the irradiance and the time of exposition; in order to monitor the exposition step, the sample were exposed at radiant exposures up to 300 kJ m$^{-2}$, depending on the exposition wavelength. The inherent ambient light irradiance of the room has been measured and is two orders of magnitude less than the light source irradiance.

Spectroscopic measurements. The UV-Visible spectroscopic measurements were performed in the total reflectance mode on a CARY5000 spectrophotometer equipped with an integrating sphere (1 s nm$^{-1}$, 1 nm step). No changes in the UV region have been observed. In order to correct small variations due to slight changes in the position of the samples, all the spectra of a series are multiplied by a correcting factor so that all the reflectance values at 260 nm are equal.

CIEL*a*b* calculation. The colorimetric coordinates were calculated in the CIE 1976 space from the reflectance spectra with the CIE formulas with a D65 illuminant and a standard photometric observer at 2° [16].

# Acknowledgement

The authors would like to thank the SACRe doctoral program (PSL University) for the financial support. This work was supported by the State managed by the National Research Agency under the Idex Sorbonne Universities as part of the Future Investments programme under the reference ANR-11-IDEX-0004-02.

# Supplementary Information

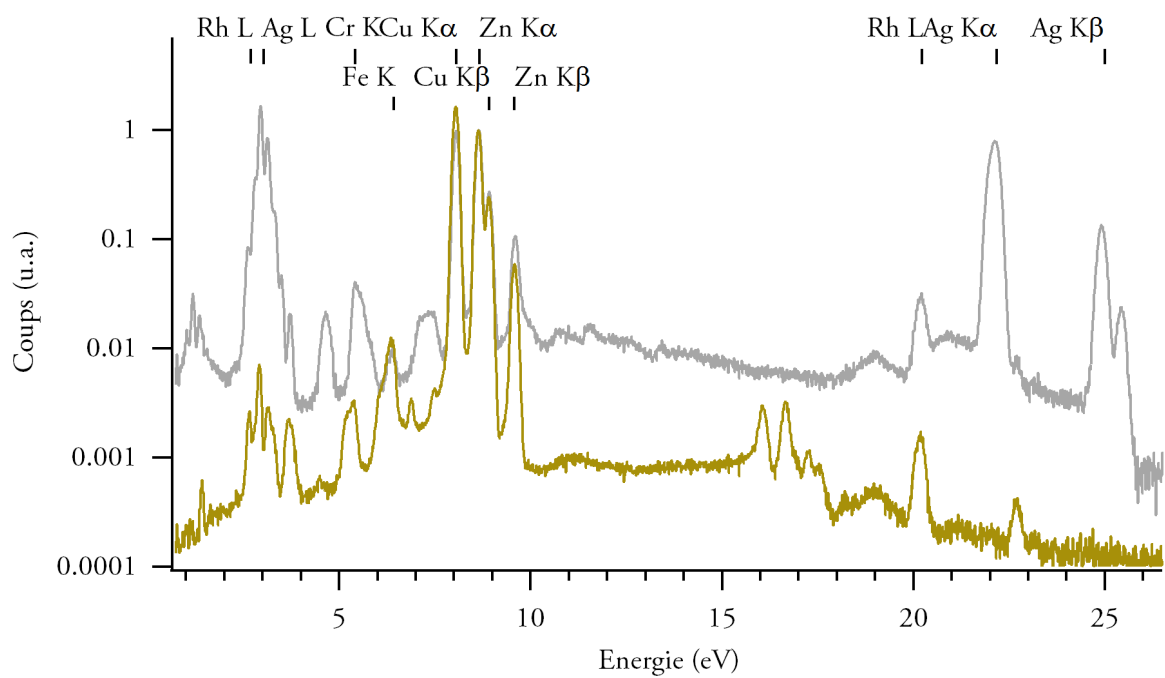

**Figure S-1.** XRF spectra of a silver plate produced by silversmith François Cadoret and gilded only on one side. The grey spectrum was acquired on the silver gilded side and the yellow one on the bare brass side. Rhodium, nickel and titanium lines are due to the X-ray source. XRF Elio XRF, Rh 40 kV, 100 mA, 480 s, 1 mm diameter beam, SDD detector. In addition to the underlying brass signal, only silver was detected on the gilded side.

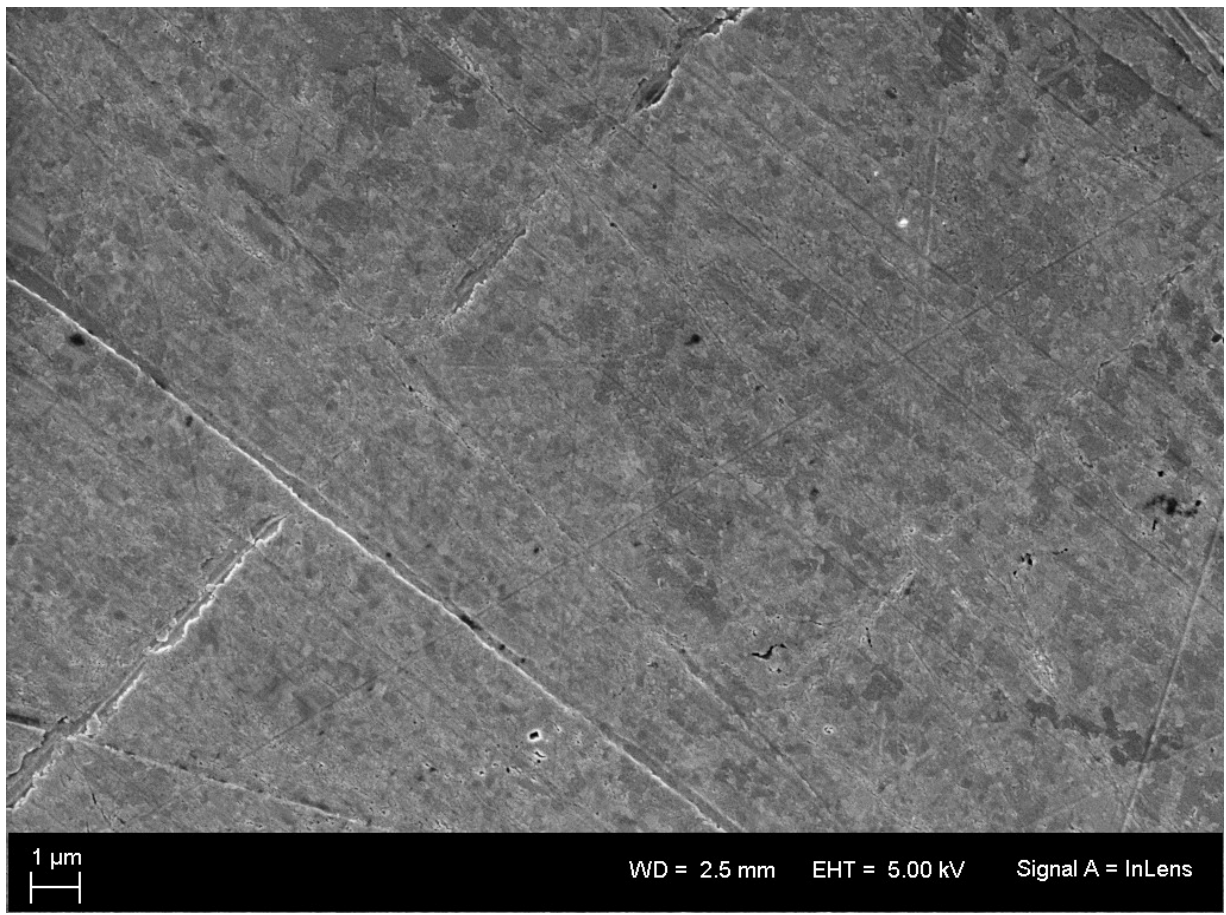

**Figure S-2.** Secondary electrons SEM image (Ultra 55 ZEISS) of a silver plate produced by silversmith François Cadoret. A few hundred nanometers large polishing trails are visible.

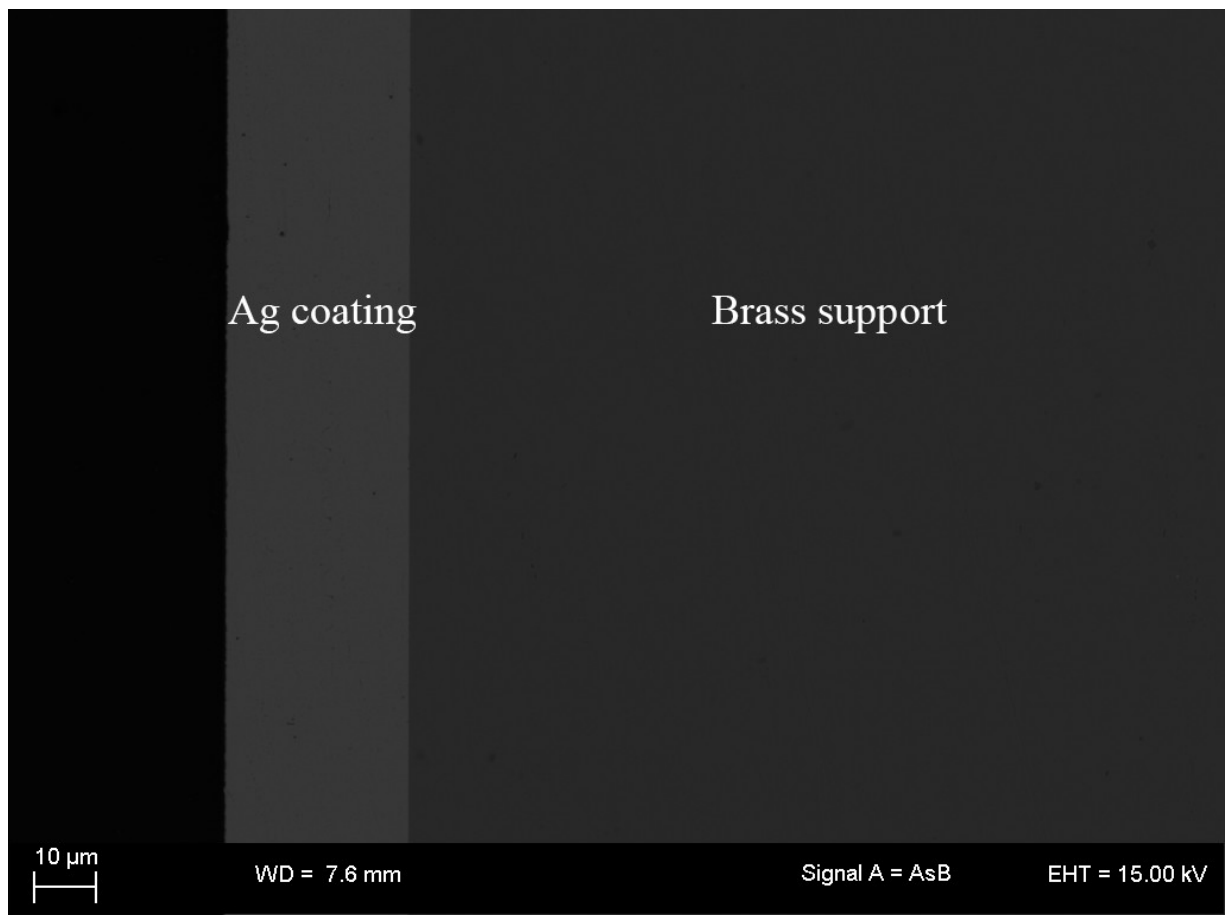

**Figure S-3.** Backscattered electrons SEM image (Ultra 55 ZEISS) of a silver plate produced by silversmith François Cadoret in cross section view. The light grey layer on the left corresponds to the silver coating of the brass support, which appears in dark grey on the right. The silver thickness was initially 50 μm and 30 μm after five successive polishing and reuse of the sample.

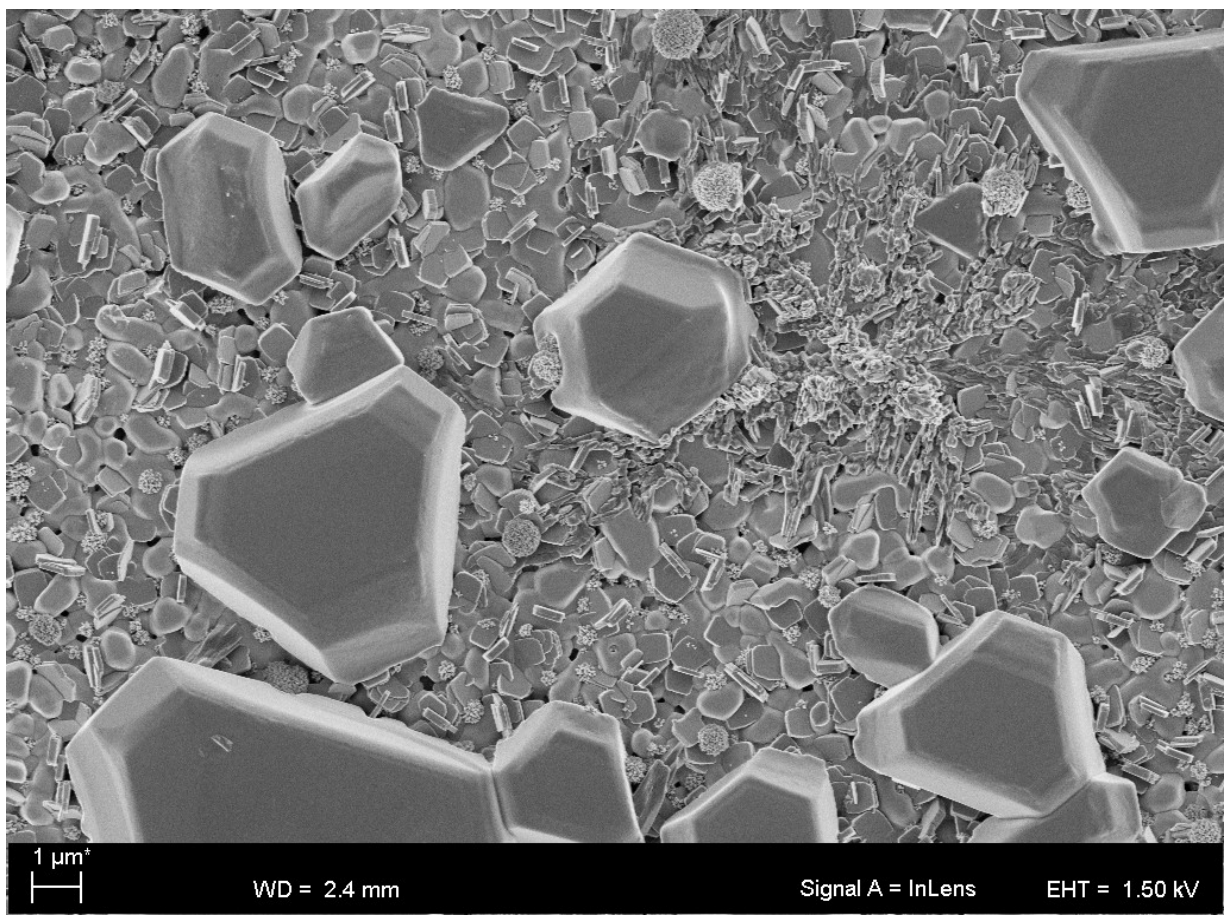

**Figure S-4.** Secondary electrons SEM image (Ultra 55 ZEISS) of an immersion-sensitized sample which has not been rinsed after the sensitization. NaCl (hexagonal) crystals are visible. EDX analysis (QUANTAX Bruker SDD) also show copper and sulfur, which can be attributed to the presence of precipitated $CuSO_4$ from the sensitization solution.

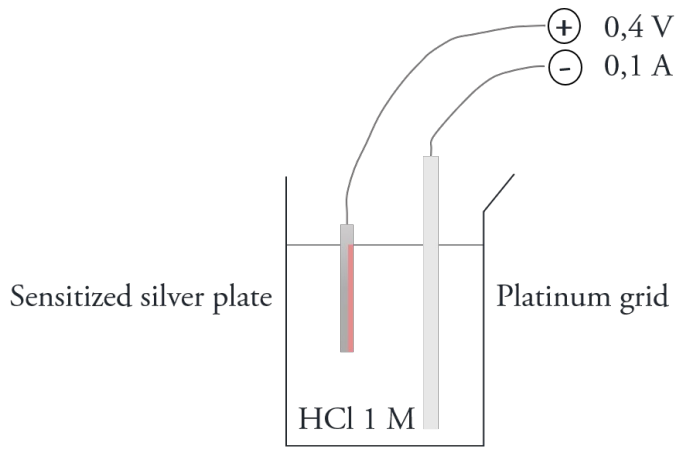

**Figure S-5.** Electrochemical sensitization: the silver plate and a platinum grid are linked respectively to the positive and negative poles of a DC generator, and soaking in a 1M HCl solution.